\documentclass[Journal]{IEEEtran}

\IEEEoverridecommandlockouts
\usepackage{color}
\usepackage{graphicx}
\usepackage{epstopdf}
\usepackage{amsmath}
\usepackage{amssymb}
\usepackage{algorithm}
\usepackage{algorithmic}
\usepackage{amsmath}
\usepackage{multirow}
\usepackage{booktabs}
\usepackage{array}
\usepackage{amsthm}
\usepackage{stfloats}
\usepackage{caption}
\usepackage{subfigure}
\usepackage{bm}
\usepackage{setspace}
\usepackage{diagbox}

\newcommand{\be}{\begin{equation}}
\newcommand{\ee}{\end{equation}}
\newcommand{\bea}{\begin{eqnarray}}
\newcommand{\eea}{\end{eqnarray}}
\newcommand{\ba}{\begin{array}}
\newcommand{\ea}{\end{array}}

\allowdisplaybreaks
\def\BibTeX{{\rm B\kern-.05em{\sc i\kern-.025em b}\kern-.08em
    T\kern-.1667em\lower.7ex\hbox{E}\kern-.125emX}}
\begin{document}

\title{Multipath Exploitation for Fluctuating Target Detection in RIS-Assisted ISAC Systems
\thanks{S. Zhang, Z. Xiao, M. Li, and W. Wang are with the School of Information and Communication Engineering, Dalian University of Technology, Dalian 116024, China (e-mail: zhangshoushuo@mail.dlut.edu.cn; xiaozichao@mail.dlut.edu.cn;  mli@dlut.edu.cn; wangwei2023@dlut.edu.cn).}
\thanks{R. Liu is with the Center for Pervasive Communications and Computing, University of California, Irvine, CA 92697, USA (e-mail: rangl2@uci.edu).}
\thanks{Q. Liu is with the School of Computer Science and Technology, Dalian University of Technology, Dalian 116024, China (e-mail: qianliu@dlut.edu.cn).}
}

\author{Shoushuo~Zhang,
        Zichao~Xiao,
        Rang~Liu,~\IEEEmembership{Member,~IEEE,}
        Ming~Li,~\IEEEmembership{Senior~Member,~IEEE,}
        Wei~Wang,~\IEEEmembership{Senior~Member,~IEEE,}
        and~Qian~Liu,~\IEEEmembership{Member,~IEEE}
}

\maketitle
\begin{abstract}
Integrated sensing and communication (ISAC)  systems are typically deployed in multipath environments, which is usually deemed as a challenging issue for wireless communications.
However, the multipath propagation can also provide extra illumination and observation perspectives for radar sensing, which offers spatial diversity gain for detecting targets with spatial radar cross section (RCS) fluctuations.
In this letter, we propose to utilize reconfigurable intelligent surfaces (RIS) in ISAC systems to provide high-quality and controllable multipath propagation for improving the performance of fluctuating target detection and simultaneously enhancing the quality of communication services.
In order to effectively exploit the spatial diversity offered by RIS-empowered multipath, the dual-functional transmit beamforming and the RIS reflection beamforming are jointly designed to maximize the expectation of radar signal-to-noise ratio (SNR).
To solve the resulting complex non-convex optimization problem, we develop an efficient alternating optimization algorithm that utilizes majorization-minimization (MM) and alternating direction method of multipliers (ADMM) algorithms.
Simulation results illustrate the advantages of multipath exploitation and the proposed beamforming design algorithm for fluctuating target detection in RIS-assisted ISAC systems.
\end{abstract}


\section{Introduction}

The emerging applications in the sixth-generation (6G) networks require a strong convergence between communication and sensing, leading to a substantial shift from simply transmitting information to actively sensing the world.
Consequently, the application requirements and the technical similarity expedite the development of integrated sensing and communication (ISAC) technology \cite{F. Liu_WC_2020}.
However, the typical urban scenarios where ISAC systems will be deployed exhibit multipath propagation, which poses a significant challenge for both wireless communication and radar sensing.

The utilization or suppression of the multipath effect is constantly debated in the fields of both communication and radar.
While a multipath environment may cause severe frequency selective fading for wireless communications, it can also provide extra perspectives of illumination and observation for radar sensing.
When considering the actual characteristics of radar cross section (RCS), the intensity of the echo will experience significant fluctuations when the observation angle changes.
Multipath can provide additional observational perspectives to compensate for RCS fading when observing from a single direction \cite{E. Fishler_TSP_2006}.
Therefore, instead of merely suppressing the multipath, researchers in radar society start the investigation on multipath exploitation to leverage the scattering diversity of the target for achieving better radar sensing performance.
In \cite{S. Sen_TSP_2011}, a generalized likelihood ratio (GLR) based method is proposed to detect an RCS fluctuating target in a multipath environment.
The authors in \cite{Z. Xu_TSP_2011} introduce a robust waveform design for a multiple input multiple output (MIMO) radar to exploit multipath.

Nevertheless, the deployment environments of ISAC systems are more intricate than radar scenarios. The multipath in such environments is characterized by its low strength, complex nature, and lack of control, which poses obstacles to the effective utilization of multipath in ISAC systems.
Fortunately, recently emerging reconfigurable intelligent surfaces (RIS) has superior capability of providing extra virtual light-of-sight (LoS) propagation paths and manipulating radio environments \cite{Y. Liu_CST_2021}-\cite{R. Liu_JSTSP_2022}.
Inspired by the advancements of RIS in communication systems, extensive research has been conducted on deploying RIS in ISAC systems.
Although the performance enhancement introduced by RIS has been validated, most of the research simplifies the fluctuation characteristics of the target  \cite{Y. Wang_ICCC_2022}, \cite{Q. Zhu_TVT_2023}.
Motivated by these findings, we propose to utilize RIS in ISAC systems to provide extra high-quality and controllable observational perspectives for fluctuating targets and simultaneously enhance communication services.

In this letter, we consider an RIS-assisted ISAC system, in which the RIS is utilized to facilitate downlink communication for multiple users as well as provide additional radar observational angle of a fluctuating target.
In order to effectively exploit the spatial diversity offered by RIS-empowered multipath, we aim to jointly design the dual-functional transmit beamforming and the RIS reflection beamforming to maximize the expectation of the signal-to-noise ratio (SNR) of the radar while satisfying communication quality-of-service (QoS), the unit-modulus constraint of RIS, and total transmit power constraints.
The majorization-minimization (MM) and alternating direction method of multipliers (ADMM) are utilized to tackle the non-convex problem.
Finally, simulation results demonstrate the advantages of RIS-empowered multipath exploitation, and conclusively show that leveraging spatial diversity can mitigate the effect of target RCS fluctuation and enhance target detection performance.

\section{System Model and Problem Formulation}
\begin{figure}[!t]
  \centering
  \vspace{-0.0 cm}
  \includegraphics[width= 2.6 in]{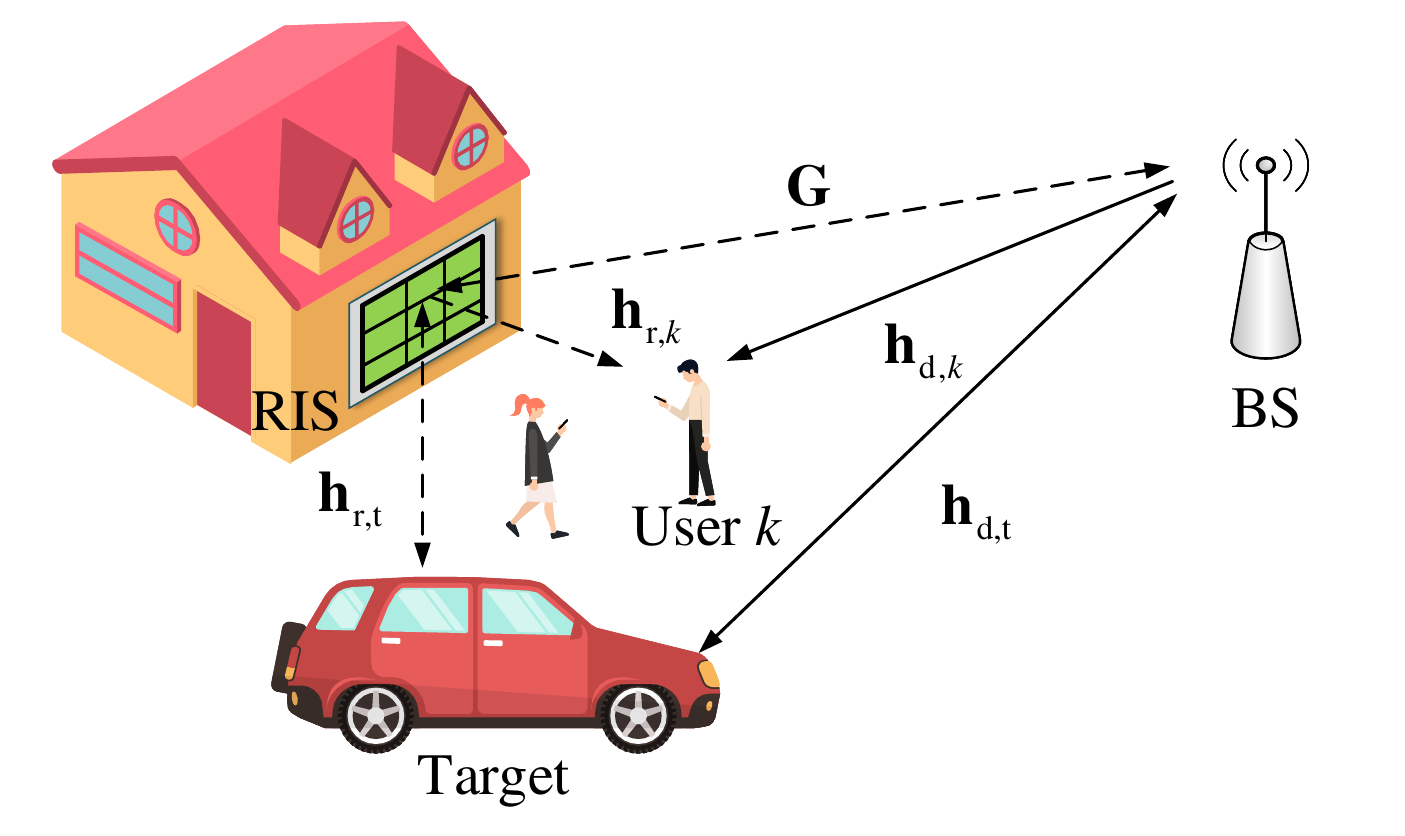}
  \caption{The considered RIS-assisted ISAC system.}
  \label{fig:systemmodel}
  \vspace{-0.5 cm}
\end{figure}

We consider a narrowband RIS-assisted ISAC system as shown in Fig. \ref{fig:systemmodel}, where a base station (BS) equipped with $M$ transmit/receive antennas detects one fluctuating target and simultaneously serves $K$ single-antenna users with the assistance of an $N$-element RIS.
Specifically, the RIS is deployed in an appropriate location, where it is near the downlink users and can provide a view different from the line of sight (LoS) to observe the target.

In order to realize fluctuating target detection, the BS transmits a dual-function pulse to illuminate the target, which has a duration of $L$ time slots.
Specifically, the transmitted signal $\mathbf{x}[l] \in \mathbb{C}^{M}$ in the $l$-th time slot can be expressed as
\begin{equation}
\mathbf{x}[l]\triangleq\mathbf{W}_\mathrm{c}\mathbf{s}_\mathrm{c}[l]+\mathbf{W}_\mathrm{r}\mathbf{s}_\mathrm{r}[l]=\mathbf{W}\mathbf{s}[l],\  \forall l,
\end{equation}
where $\mathbf{W}_\mathrm{c} \triangleq [\mathbf{w}_1,\ldots,\mathbf{w}_K] \in \mathbb{C}^{M\times K}$ and $\mathbf{W}_\mathrm{r}\in \mathbb{C}^{M\times M}$ denote the communication beamforming matrix and the radar beamforming matrix, respectively.
The vector $\mathbf{s}_\mathrm{c}[l] \in \mathbb{C}^K$ represents the communication symbols of the $K$ users with $\mathbb{E}\{\mathbf{s}_\mathrm{c}[l]\mathbf{s}_\mathrm{c}[l]^H\}=\mathbf{I}_K$.
The vector $\mathbf{s}_\mathrm{r}[l] \in \mathbb{C}^M$ denotes the radar probing signals with $\mathbb{E}\{\mathbf{s}_\mathrm{r}[l]\mathbf{s}_\mathrm{r}[l]^H\}=\mathbf{I}_M$, which is statistically independent with $\mathbf{s}_\mathrm{c}[l]$.
For brevity, we define the overall beamforming matrix as $\mathbf{W}\triangleq [\mathbf{W}_\text{c}, \mathbf{W}_\text{r}] \in \mathbb{C}^{M \times (K+M)}$  and the transmit symbols as $\mathbf{s}[l]\triangleq [\mathbf{s}_{\mathrm{c}}[l]^T, \mathbf{s}_{\mathrm{r}}[l]^T]^T\in \mathbb{C}^{K+M},\ \forall l$.

The target echo signal received by the BS includes the components of four different paths as shown in Fig. \ref{fig:systemmodel}.
Due to severe path loss, the signal that propagates through RIS twice can be neglected.
Therefore, the baseband echo signal received at the BS in the $l$-th time slot can be written as
\begin{align}
\mathbf{y}_\mathrm{r}[l]=&\ \alpha_0 \mathbf{h}_{\mathrm{d,t}}\mathbf{h}_{\mathrm{d,t}}^T\mathbf{x}[l]+\alpha_1(\mathbf{h}_{\mathrm{d,t}}\mathbf{h}_{\mathrm{r,t}}^T\mathbf{\Phi}\mathbf{G}+\mathbf{G}^T\mathbf{\Phi}\mathbf{h}_{\mathrm{r,t}}\mathbf{h}_{\mathrm{d,t}}^T)\nonumber\\
&\ \mathbf{x}[l-\tau]+\mathbf{z}[l]  \label{echo signal}\\
=&\ \alpha_0 \mathbf{H}_0\mathbf{Ws}[l]+\alpha_1\mathbf{H}_1(\bm\phi)\mathbf{Ws}[l-\tau]+\mathbf{z}[l], \ \forall l, \nonumber
\end{align}
where $\alpha_{0}$ and $\alpha_1$ denote the fluctuating target RCS coefficients towards BS and RIS, respectively, and are considered as classic Swerling model \uppercase\expandafter{\romannumeral1}.
Correspondingly, $\alpha_{0}$ and $\alpha_1$ can be modeled as independent complex Gaussian random variables, i.e., $\alpha_0 \sim \mathcal{CN}(0,\sigma^2_0)$, $\alpha_1 \sim \mathcal{CN}(0,\sigma^2_1)$.
In (\ref{echo signal}), the vectors $\mathbf{h}_{\mathrm{d,t}} \in \mathbb{C}^M$ and $\mathbf{h}_{\mathrm{r,t}} \in \mathbb{C}^N$ respectively denote the channels between the BS and the target and between the RIS and the target, which are generally LoS.
The matrix $\mathbf{G} \in \mathbb{C}^{M \times N}$ represents the channel between the BS and the RIS.
The RIS reflection matrix is defined by $\bm{\Phi}\triangleq \mathrm{diag}\{\bm{\phi}\}$ with $\bm{\phi}=[\phi_1,\phi_2,\ldots,\phi_N]^T$.
The constant $\tau$ stands for the relative delay in the fast time domain, which can be predicted by geometric information or ray-tracing \cite{H. T. Hayvaci_RSN_2013}.
The vector $\mathbf{z}[l] \in \mathbb{C}^M$ denotes the additive white Gaussian noise (AWGN) and $\mathbf{z}[l] \sim \mathcal{CN}(\mathbf{0},\sigma_\mathrm{z}^2 \mathbf{I}_M)$.
For simplicity, with $\mathbf{h}_{\mathrm{r,t}}^T\bm{\Phi}=\bm{\phi}^T\mathrm{diag}\{\mathbf{h}_{\mathrm{r,t}}\}$, we define
\begin{subequations}
\begin{align}
\mathbf{H}_0 &\triangleq \mathbf{h}_{\mathrm{d,t}}\mathbf{h}_{\mathrm{d,t}}^T,\\
\mathbf{H}_1(\bm{\phi}) & \triangleq \mathbf{h}_{\mathrm{d,t}}\bm{\phi}^T\text{diag}\{\mathbf{h}_{\mathrm{r,t}}\}\mathbf{G}+\mathbf{G}^T\text{diag}\{\mathbf{h}_{\mathrm{r,t}}\}\bm{\phi}\mathbf{h}_{\mathrm{d,t}}^T .
\end{align}
\end{subequations}

After collecting $Q= L+\tau$ snapshots, all the received signals to be processed  in a matrix form as
\begin{equation}
\begin{aligned}
    \mathbf{Y}_\text{r} &\triangleq [\mathbf{y}_\mathrm{r}[1],\mathbf{y}_\mathrm{r}[2],\ldots,\mathbf{y}_\mathrm{r}[Q]]\\
    &= \alpha_0 \mathbf{H}_0\mathbf{WSJ}_0+\alpha_1\mathbf{H}_1(\bm\phi)\mathbf{WSJ}_1+\mathbf{Z},
\end{aligned}
\end{equation}
where $\mathbf{S}\triangleq [\mathbf{s}[1],\mathbf{s}[2],\ldots,\mathbf{s}[L]]$ and $\mathbf{Z}\triangleq [\mathbf{z}[1],\mathbf{z}[2],\ldots,\mathbf{z}[Q]]$.
In order to appropriately represent the propagation delay, we introduce two shift matrices $\mathbf{J}_0 \in \mathbb{C}^{L\times Q} $ and $\mathbf{J}_1  \in \mathbb{C}^{L\times Q} $, which are respectively defined as $\mathbf{J}_0(m,n) \triangleq \begin{cases} 1, & n-m=0 \\ 0, & \text{otherwise} \end{cases}  $  and $\mathbf{J}_1(m,n) \triangleq \begin{cases} 1, & n-m= \tau \\ 0, & \text{otherwise} \end{cases}$.
In order to facilitate the following signal processing, we further vectorize the received signals $\mathbf{Y}_\text{r}$ as follows
\begin{subequations}
\begin{align}
\mathbf{y}_\mathrm{r}&\triangleq \text{vec}\{\mathbf{Y}_\text{r}\}\\
&=\left[\alpha_0\mathbf{J}_0^T\otimes\mathbf{H}_0+\alpha_1\mathbf{J}_1^T\otimes\mathbf{H}_1(\bm{\phi})\right](\mathbf{I}_L\otimes \mathbf{W})\mathbf{s}+\mathbf{z}\\
& = [\alpha_0\widetilde{\mathbf{H}}_0+\alpha_1\widetilde{\mathbf{H}}_1(\bm{\phi})]\widetilde{\mathbf{W}}\mathbf{s}+\mathbf{z},
\end{align}
\end{subequations}
where $\widetilde{\mathbf{H}}_0 \triangleq  \mathbf{J}_0^T\otimes\mathbf{H}_0$, $\widetilde{\mathbf{H}}_1 \triangleq  \mathbf{J}_1^T\otimes\mathbf{H}_1(\bm{\phi})$, $\widetilde{\mathbf{W}} \triangleq \mathbf{I}_L\otimes\mathbf{W}$, $\mathbf{s}\triangleq\text{vec}\{\mathbf{S}\}$, and $\mathbf{z}\triangleq\text{vec}\{\mathbf{Z}\}$.

To distinguish the multipath echo signals, joint space-time processing for the received signals is proposed \cite{R. Liu_JSTSP_2022}.
After applying the space-time receive filter $\mathbf{f} \in \mathbb{C}^{MQ}$ to process the received signal $\mathbf{y}_\text{r}$, the output $\mathbf{f}^H\mathbf{y}_\mathrm{r}$ is used for Neyman-Pearson detector to determine the presence or absence of the target.
Since the output SNR is positively correlated with the detection probability, it will be used as the sensing metric, which is calculated as
\begin{subequations}
\begin{align}
\label{instantaneous SNR}
    \text{SNR}=&\frac{\mathbb{E}\{|\mathbf{f}^H(\alpha_0\widetilde{\mathbf{H}}_0+\alpha_1\widetilde{\mathbf{H}}_1)\widetilde{\mathbf{W}}\mathbf{s}|^2\}}{\sigma_\mathrm{z}^2\mathbf{f}^H\mathbf{f}}\\
\label{expectation of SNR}
=&\frac{|\mathbf{f}^H\sigma_0\widetilde{\mathbf{H}}_0\widetilde{\mathbf{W}}\mathbf{s}|^2+|\mathbf{f}^H\sigma_1\widetilde{\mathbf{H}}_1\widetilde{\mathbf{W}}\mathbf{s}|^2}{\sigma_\mathrm{z}^2\mathbf{f}^H\mathbf{f}}.
\end{align}
\end{subequations}
Due to the fluctuating RCS of two different paths, the form of radar SNR is not conducive to solving the filter, and thus some algebraic transformations are necessary.
Since the symbols from different time slots are statistically independent (i.e. $\mathbb{E}\{\mathbf{s}[l]\mathbf{s}[l+\tau]^H\}=\mathbf{0}_{K+M}$),
we can prove the orthogonality of the signal components from different paths
\begin{equation}
\begin{split}
&\mathbb{E}\{\mathbf{s}^H\widetilde{\mathbf{W}}^H\widetilde{\mathbf{H}}_0^H\widetilde{\mathbf{H}}_1\widetilde{\mathbf{W}}\mathbf{s}\}\\
=&\Sigma_{l=1}^{L-\tau}\mathbb{E}\{\mathbf{s}[l]^H \mathbf{W}^H \mathbf{H}_0^H \mathbf{H}_1(\bm{\phi})\mathbf{W}\mathbf{s}[l+\tau]\}\\
=&0.
\end{split}
\end{equation}
Then, based on this orthogonality, the SNR maximization problem with respect to $\mathbf{f}$ can be converted to a Rayleigh quotient problem as
\begin{equation}\label{Rayleigh quotient}
\max_{\mathbf{f}} \frac{\mathbf{f}^H\mathbf{A}\mathbf{f}}{\sigma_\mathrm{z}^2\mathbf{f}^H\mathbf{f}},
\end{equation}
where we define
\begin{equation}
\mathbf{A} \triangleq(\sigma_0\widetilde{\mathbf{H}}_0\widetilde{\mathbf{W}}\mathbf{s}+\sigma_1\widetilde{\mathbf{H}}_1\widetilde{\mathbf{W}}\mathbf{s})(\sigma_0\widetilde{\mathbf{H}}_0\widetilde{\mathbf{W}}\mathbf{s}+\sigma_1\widetilde{\mathbf{H}}_1\widetilde{\mathbf{W}}\mathbf{s})^H.
\end{equation}
Hence, the optimal solution for \eqref{Rayleigh quotient} is attained as
\begin{equation}\label{filter}
\mathbf{f}^\star=\frac{(\sigma_0\widetilde{\mathbf{H}}_0+\sigma_1\widetilde{\mathbf{H}}_1)\widetilde{\mathbf{W}}\mathbf{s}}{\|(\sigma_0\widetilde{\mathbf{H}}_0+\sigma_1\widetilde{\mathbf{H}}_1)\widetilde{\mathbf{W}}\mathbf{s}\|_2}.
\end{equation}
Substituting \eqref{filter}, the radar SNR is reformulated as
\begin{equation}
\text{SNR}=\frac{L}{\sigma_\mathrm{z}^2}\mathrm{Tr}\{\mathbf{W}^H(\sigma_0^2\mathbf{H}_0^H\mathbf{H}_0+\sigma_1^2\mathbf{H}_1^H(\bm{\phi})\mathbf{H}_1(\bm{\phi}))\mathbf{W}\}.
\end{equation}

From the communication perspective, the received signal at the $k$-th user in the $l$-th time slot can be expressed as
\begin{equation}
r_k[l]=(\mathbf{h}_{\mathrm{d},k}^T+\mathbf{h}_{\mathrm{r},k}^T\bm{\Phi}\mathbf{G})\mathbf{x}[l]+n_k[l],\ \forall l,\ \forall k,
\end{equation}
where $\mathbf{h}_{\text{d},k}\in \mathbb{C}^M$ and $\mathbf{h}_{\text{r},k} \in \mathbb{C}^N$ denote the channels between the BS and the $k$-th user, and between the RIS and the $k$-th user, respectively.
$n_k[l]$ stands for AWGN of the $k$-th user and $n_k[l]\sim \mathcal{CN}(0,\sigma_k^2)$.
Accordingly, the communication signal-to-interference-plus-noise ratio (SINR) of the $k$-th user can be given as
\vspace{-0.1 cm}
\begin{equation}
\gamma_k=\frac{|(\mathbf{h}_{\text{d},k}^T+\mathbf{h}_{\text{r},k}^T\bm{\Phi}\mathbf{G})\mathbf{w}_k|^2}{\sum_{j\neq k}^{K+M}|(\mathbf{h}_{\text{d},k}^T+\mathbf{h}_{\text{r},k}^T\bm{\Phi}\mathbf{G})\mathbf{w}_j|^2+\sigma_k^2},\ \forall k.
\end{equation}

In this paper, we aim to maximize the radar SNR while satisfying the communication QoS requirement, the total transmit power budget, and the unit-modulus phase-shift of the RIS by the joint beamforming $\mathbf{W}$ and passive reflection $\bm{\phi}$ design.
The optimization problem can be formulated as
\begin{subequations}
\label{eq:obj}
\begin{align}
    \label{eq:obj_a}
    \max_{\mathbf{W}, \bm{\phi}}\ &\ \text{SNR}(\mathbf{W},\bm{\phi})\\
    \label{eq:obj_b}
    \mathrm{s.t.}\ \ &\  \gamma_k\geq \Gamma_k,\ \forall k,\\
    \label{eq:obj_c} &\ \|\mathbf{W}\|_F^2\leq P,\\
    \label{eq:obj_d} &\ |\phi_n|=1,\ \forall n,
\end{align}
\end{subequations}
where $\Gamma_k$ denotes the $k$-th user communication SINR requirement, and $P$ is the total power budget.
It is obvious that the problem \eqref{eq:obj} is a complicated non-convex problem, which is difficult to optimize the non-convex objective function \eqref{eq:obj_a}, the unit-modulus constraint \eqref{eq:obj_d}, and the mutually coupled variables $\mathbf{W}$ and $\bm{\phi}$.
In order to tackle these difficulties, an efficient alternating optimization algorithm is proposed in the next section.

\vspace{-0.0 cm}
\section{Joint Beamforming and RIS Reflection Design}

\subsection{MM-Based Transformation}
In order to handle the non-convex quadratic objective function \eqref{eq:obj_a} and constraints \eqref{eq:obj_b}, we propose to utilize the MM method to construct a series of more tractable surrogate functions.
Specifically, with the given local points $\mathbf{W}^{(t)}$ and $\bm{\phi}^{(t)}$ in the $t$-th iteration, the lower bound for the objective function is given by
\begin{align}
&\mathrm{Tr}\{\mathbf{W}^H(\sigma_0^2\mathbf{H}_0^H\mathbf{H}_0+\sigma_1^2\mathbf{H}_1^H(\bm{\phi})\mathbf{H}_1(\bm{\phi}))\mathbf{W}\}\geq 2\Re\{\mathrm{Tr}\{\nonumber\\
&(\mathbf{W}^{(t)})^H(\sigma_0^2\mathbf{H}_0^H\mathbf{H}_0+\sigma_1^2\mathbf{H}_1^H(\bm{\phi}^{(t)})\mathbf{H}_1(\bm{\phi}))\mathbf{W}\}\}-c_1^{(t)}, \label{eq:surrogate function}
\end{align}
where for conciseness, we define
\begin{align}    c_1^{(t)}=\sigma_0^2\|\mathbf{H}_0\mathbf{W}^{(t)}\|_F^2+\sigma_1^2\|\mathbf{H}_1(\bm{\phi}^{(t)})\mathbf{W}^{(t)}\|_F^2.
\end{align}

With the surrogate function (\ref{eq:surrogate function}), we define a new objective function $f(\mathbf{W},\bm{\phi})$ to facilitate the algorithm development, which is a linear function with respect to $\mathbf{W}$ and $\bm{\phi}$ as
\begin{subequations}\begin{align}
    f(\mathbf{W},\bm{\phi})&\triangleq
\Re\{\mathrm{Tr}\{\mathbf{F}_1^{(t)}\mathbf{W}+\mathbf{F}_2^{(t)}\mathbf{H}_1(\bm{\phi})\mathbf{W}\}\} ,  \\
 \mathbf{F}_1^{(t)}&\triangleq\sigma_0^2(\mathbf{W}^{(t)})^H\mathbf{H}_0^H\mathbf{H}_0 ,\\
 \mathbf{F}_2^{(t)}&\triangleq\sigma_1^2(\mathbf{W}^{(t)})^H\mathbf{H}_1^H(\bm{\phi}^{(t)}).
\end{align}\end{subequations}

For the non-convex constraints \eqref{eq:obj_b}, after some algebra transformations, they can be reformulated as
\begin{equation}\label{constraint b}\Gamma_k^{-1}|\mathbf{h}_k^T(\bm{\phi})\mathbf{w}_k|^2-\sum_{j\neq k}^{K+M}|\mathbf{h}_k^T(\bm{\phi})\mathbf{w}_j|^2-\sigma_k^2 \geq 0,\ \forall k,
\end{equation}
where $\mathbf{h}_k(\bm{\phi})\triangleq \mathbf{h}_{\text{d},k}+\mathbf{G}^T\mathrm{diag}\{\mathbf{h}_{\text{r},k}\}\bm{\phi},\ \forall k$.
Since the first quadratic term in \eqref{constraint b} makes the constraint non-convex, we attempt to utilize the MM method to relax it by
\begin{align}\label{relaxtion}
|\mathbf{h}_k^T(\bm{\phi})\mathbf{w}_k|^2\geq 2\Re\{\mathbf{h}_k^T(\bm{\phi}^{(t)})\mathbf{w}_k^{(t)}\mathbf{h}^H(\bm{\phi})\mathbf{w}_k^\ast\}-c_2^{(t)},\ \forall k,
\end{align}
where $c_2^{(t)}\triangleq |\mathbf{h}_k^T(\bm{\phi}^{(t)})\mathbf{w}_k^{(t)}|^2$.
Then, substituting the relaxation \eqref{relaxtion} into constraints \eqref{constraint b}, their surrogates $g_k(\mathbf{W},\bm{\phi})$ are constructed as
\begin{equation}
    c_3^{(t)}\Re\{\mathbf{h}_k^H(\bm{\phi})\mathbf{w}_k^{\ast}\} - \sum_{j\neq k}^{K+M}|\mathbf{h}_k^T(\bm{\phi})\mathbf{w}_j|^2-c_4^{(t)} \geq 0,~ \forall k,
\end{equation}
which are convex with respect to $\mathbf{w}$ and $\bm{\phi}$, respectively.
For simplicity, we define
\begin{subequations}
    \begin{align}
        c_3^{(t)}&\triangleq2\Gamma_k^{-1}\Re\{\mathbf{h}_k^T(\bm{\phi}^{(t)})\mathbf{w}_k^{(t)}\},\ \forall k,\\
     c_4^{(t)}&\triangleq  \Gamma_k^{-1}c_2^{(t)}+\sigma_k^2,\ \forall k.
    \end{align}
\end{subequations}

\subsection{ADMM-Based Transformation}
After solving the non-convex objective function \eqref{eq:obj_a} and constraints \eqref{eq:obj_b}, only the constant modulus constraints \eqref{eq:obj_d} of RIS hinder the optimization of problem \eqref{eq:obj}.
For the sake of dealing with the non-convex constant-modulus constraint, we first introduce the auxiliary variable $\bm{\psi}=\bm{\phi}$ to transform it into a separable form and then utilize the ADMM method to decouple it into a simple sub-problem with an analytic solution.
Specifically, problem \eqref{eq:obj} is first converted to
\begin{subequations}
\label{eq:obj2}
\begin{align}
    \label{eq:obj2_a}
    \max_{\mathbf{W}, \bm{\phi}, \bm{\psi}}&\ f(\mathbf{W},\bm{\phi})\\
    \label{eq:obj2_b}
    \mathrm{s.t.}\ &\ g_k(\mathbf{W},\bm{\phi})\geq 0,\ \forall k,\\
    \label{eq:obj2_c} &\ \|\mathbf{W}\|_F^2\leq P,\\
    \label{eq:obj2_d} &\ |\phi_n|\leq1, \forall n,\\
    \label{eq:obj2_e} &\ |\psi_n|=1, \forall n,\\
    \label{eq:obj2_f} &\ \bm{\psi}=\bm{\phi}.
\end{align}
\end{subequations}

This problem can be efficiently solved in an ADMM manner, which
alternately updates the variables in maximizing its augmented Lagrangian (AL) function.
In specific, the AL function of \eqref{eq:obj2} is constructed as
\begin{equation}
L_{\bm{\lambda}}(\mathbf{W},\bm{\phi},\bm{\psi})=f(\mathbf{W},\bm{\phi})- \frac{\rho}{2}\|\bm{\phi}-\bm{\psi}-\rho^{-1}\bm{\lambda}\|^2,
\end{equation}
where $\bm{\lambda}$ is the dual variable, and $\rho > 0$ is the penalty parameter.

\subsection{Block Update}
1) \textit{Update} $\mathbf{W}$: with given $\bm{\phi}$, $\bm{\psi}$, and $\bm{\lambda}$ and ignoring the constant term, the optimization with respect to $\mathbf{W}$ is expressed as
\begin{subequations}
\label{eq:obj4}
\begin{align}
    \label{eq:obj4_a}
    \max_{\mathbf{W}}&\ f(\mathbf{W}, \bm{\phi}^{(t)}) \\
    \label{eq:obj4_b}
    \mathrm{s.t.}&\ g_k(\mathbf{W}, \bm{\phi}^{(t)}) \geq 0,\ \forall k,\\
    \label{eq:obj4_c} &\ \|\mathbf{W}\|_F^2\leq P,
\end{align}
\end{subequations}
which is obviously convex with respect to $\mathbf{W}$ and can be effectively solved by CVX.

2) \textit{Update} $\bm{\phi}$: with given $\mathbf{W}$, $\bm{\psi}$, and $\bm{\lambda}$, the optimization problem with respect to $\bm{\phi}$ can be expressed as
\begin{subequations}
\label{eq:obj5}
\begin{align}
    \label{eq:obj5_a}
    \max_{\bm{\phi}}&\ L_{\bm{\lambda}}(\mathbf{W}^{(t)},\bm{\phi},\bm{\psi}) \\
    \label{eq:obj5_b}
    \mathrm{s.t.}&\ g_k(\mathbf{W}^{(t)}, \bm{\phi})\geq 0,\ \forall k,\\
    \label{eq:obj5_c} &\ |\phi_n|\leq1,\ \forall n.
\end{align}
\end{subequations}
It is obvious that it is a convex problem, which can also be effectively solved by CVX.

3) \textit{Update} $\bm{\psi}$: when $\bm{\phi}$ is fixed, $\bm{\psi}$ is updated by solving the following problem:
\begin{subequations}\begin{align}\label{uadate psi}
\min _{\bm{\psi}}&\  \frac{\rho}{2}\|\bm{\phi}-\bm{\psi}+\rho^{-1}\bm{\lambda}\|_2^2,\\
&\text{s.t.} \ |\psi_n|=1, \forall n.
\end{align}\end{subequations}
Note that the value of the quadratic term of the objective function \eqref{uadate psi} remains constant when $\bm{\psi}$ has unit modulus entries.
The maximum value of the objective function \eqref{uadate psi} is attained $\bm{\psi}$ is aligned with the linear part $\rho\bm{\phi}+\bm{\lambda}$ as
\begin{equation}\label{eq34}
\bm{\psi}=e^{j\angle(\rho\bm{\phi}+\bm{\lambda})}.
\end{equation}

4) \textit{Update} $\bm{\lambda}$: after updating $\bm{\phi}$ and $\bm{\psi}$, the dual variable $\bm{\lambda}$ is updated by
\begin{equation}\label{eq35}
\bm{\lambda}:=\bm{\lambda}+\rho(\bm{\phi}-\bm{\psi}).
\end{equation}

\vspace{-0.4 cm}

\subsection{Initialization and Complexity Analysis}
For the proposed alternating algorithm,
we initialize $\bm{\phi}$ by maximizing the channel gains of the target and the users, subject to the constant-modulus constraint of RIS.
Specifically, the optimization problem for initializing $\bm{\phi}$ is formulated as
\begin{subequations}\label{initialization}
\begin{align}
\label{initialization a}
\max _{\bm{\phi}} &\ \|\mathbf{H}_1(\bm{\phi})\|_F^2+\sum_{k=1}^{K}\|\mathbf{h}_k(\bm{\phi})\|^2\\
\label{initialization b}
\mathrm{s.t.} & |\phi_n|=1, \forall n,
\end{align}
\end{subequations}
whose objective function of \eqref{initialization} is smooth and differentiable.
Additionally, the unit-modulus constraint \eqref{initialization b} forms a complex circle Riemannian manifold, which allows problem \eqref{initialization} to be effectively solved by the Riemannian algorithm \cite{R. Liu_JSTSP_2022}.

In the initialization stage, the complexity for obtaining $\bm{\phi}$ by the Riemannian optimization is of order $\mathcal{O}\{N^{1.5}\}$.
The complexity for updating $\mathbf{W}$ by solving \eqref{eq:obj4} is order $\mathcal{O}\{\sqrt{K+1}M^3(K+M)^3\}$.
Solving problem \eqref{eq:obj5} to update $\mathbf{\bm{\phi}}$ has the complexity of order $\mathcal{O}\{\sqrt{K+1}N^3\}$.
Updating $\bm{\psi}$ and $\bm{\lambda}$ both have the  order $\mathcal{O}\{N\}$.
Therefore, the overall complexity of the proposed algorithm is of order $\mathcal{O}\{\sqrt{K+1}[M^3(K+M)^3+N^3]\}$.

\section{Simulation Results}

\begin{figure}[t]
  \centering
      \vspace{-0.5 cm}
  \includegraphics[width=  2.6  in]{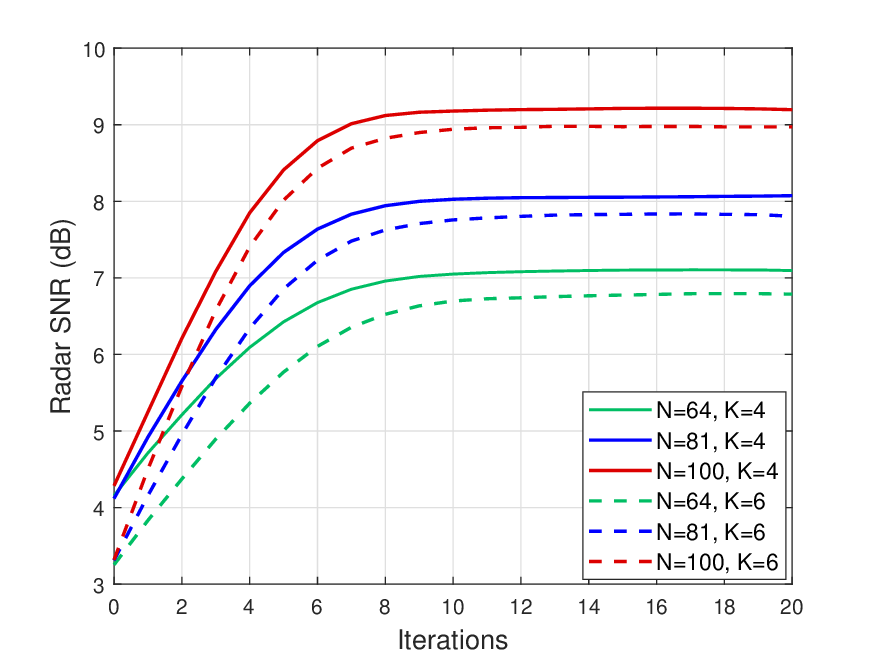}
  \caption{Convergence performance ($P=30\mathrm{W},\Gamma_k=10\mathrm{dB}$).}\label{fig:fig2}
   \vspace{-0.6 cm}
\end{figure}

In this section, numerical results are presented to show the advantages of effectively exploiting the controllable multipath for the fluctuating target detection in the RIS-empowered ISAC system.
It is assumed that the BS equipped with $M = 16$ antennas in half-wavelength spacing detects a fluctuating target and simultaneously serves $K$ single antenna users with the assistance of an $N$ element RIS.
The length of the transmit dual-function signal is set as $L = 64$, and the relative delay $\tau$ is 16 samples.
The distances of BS-target, BS-RIS, BS-users, RIS-target, and RIS-user links are set as 50m, 40m, 36m, 25m, and 3m, respectively.
We adopt the typical path-loss model and set the path-loss exponents of BS-RIS, RIS-user, and BS-user links as $\alpha_\mathrm{BR}=2.0$, $\alpha_\mathrm{RU}=2.4$, and $\alpha_\mathrm{{BU}}=2.7$.
The Rician fading channel model is assumed with the Rician factors of BS-RIS, BS-target link being $\beta_\mathrm{BR}=5$ dB and $\beta_\mathrm{Bt}\rightarrow \infty$ respectively, and other links being 0dB.
The variances of RCS $\alpha_0$ and $\alpha_1$ are assumed to be $\sigma_0^2=\sigma_1^2=1$.
The noise power at the radar receiver and the $k$-th user is set as $\sigma_\mathrm{z}^2=\sigma_k^2=-80$ dBm.

In Fig.~\ref{fig:fig2}, we first show the convergence performance of the proposed joint beamforming design algorithm. It can be clearly seen that the radar SNRs under different settings converge to a satisfactory level within 20 iterations, which verifies the efficiency and effectiveness of the proposed algorithm.

\begin{figure}[t]
  \centering
      \vspace{-0.0 cm}
  \includegraphics[width= 2.8 in]{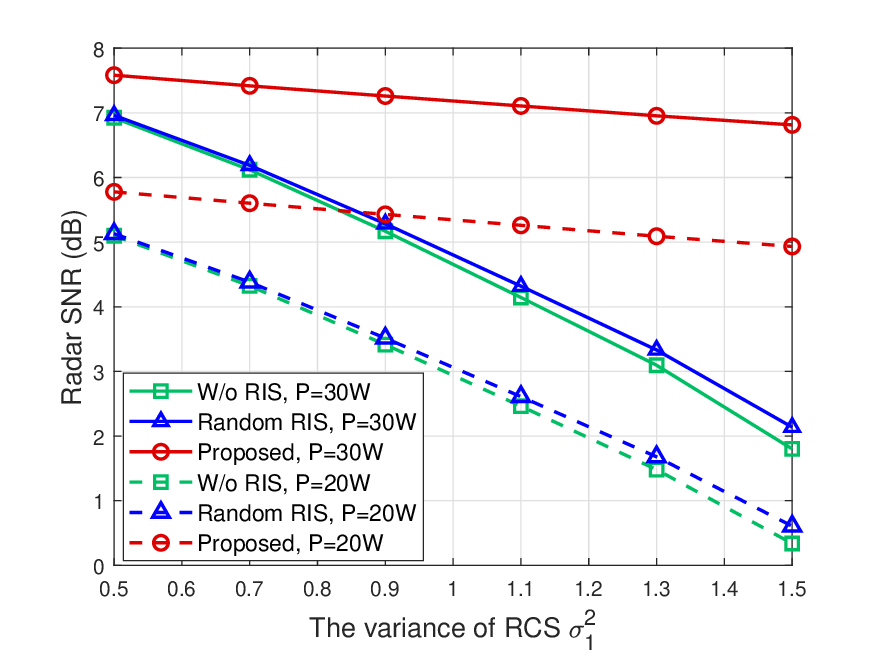}
  \caption{Radar SNR versus the variance of RCS $\sigma_1^2$ ($K=4, N=64, \Gamma_k=10\mathrm{dB}$).}\label{fig:fig3}
  \vspace{-0.5 cm}
\end{figure}

In order to clearly demonstrate the advantages of multipath exploitation, in Fig.~\ref{fig:fig3} we show the radar SNR versus the variance of RCS coefficient $\sigma_1^2$ of the RIS path with a fixed summation of RCS variances, i.e., $\sigma_0^2 + \sigma_1^2 = 2$.
The schemes without RIS (``W/o RIS''), and random phase-shift RIS (``Random RIS'') are included for comparison purposes. The solid lines and dashed lines represent the total power budgets of 30W and 20W, respectively.
It can be noticed that, with the increase of $\sigma_1^2$, the radar SNR of our proposed joint beamforming design algorithm maintains a relatively high level while that of the counterparts greatly decreases due to the smaller target RCS towards the BS.
This phenomenon verifies that exploiting the extra observation perspectives provided by the optimized RIS can enrich spatial diversity and significantly enhance the detection performance when the RCS of the target fluctuates.

\begin{figure}[t]
  \centering
  \includegraphics[width= 2.8  in]{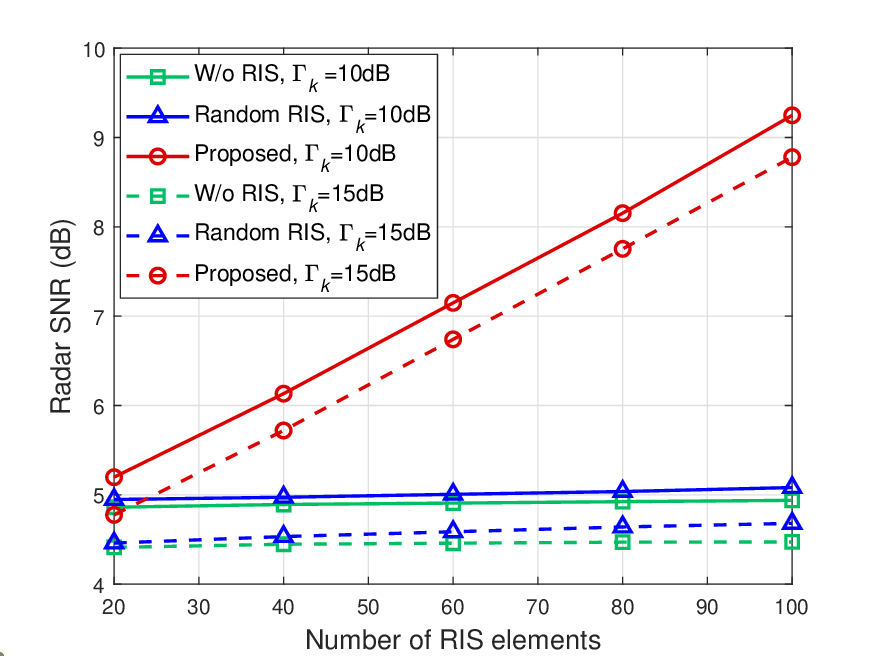}
  \caption{Radar SNR versus the number of elements $N$ ($K=4, P=30\mathrm{W}$).}\label{fig:fig4}
   \vspace{-0.4 cm}
 \end{figure}
Next, the radar SNR versus the number of RIS elements is illustrated in Fig.~\ref{fig:fig4}.
With our proposed joint beamforming and reflection design, the radar SNR obviously increases with the increasing number of RIS elements in the considered RIS-assisted ISAC system.
Additionally, with the increasing demand for communication QoS, the radar SNR will suffer a certain loss due to the performance trade-off between communication and radar functions.

 \begin{figure}[t]
 \centering
   \vspace{-0.0 cm}
  \includegraphics[width=  2.8  in]{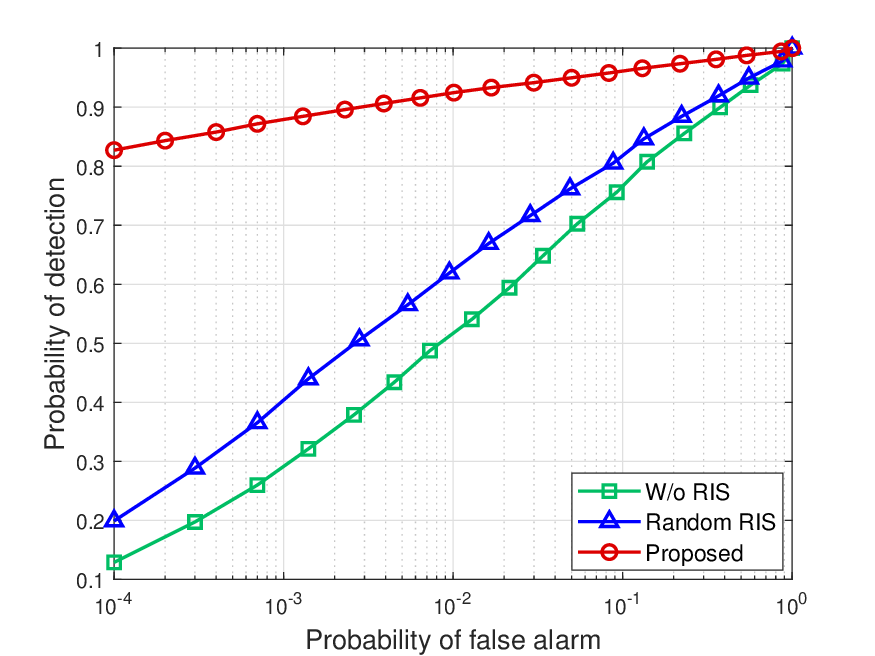}
  \caption{Probability of detection as a function of the probability of false alarm ($K=4, P=30 \mathrm{W}, N=64, \Gamma_k=10 \mathrm{dB}$).}\label{fig:fig5}
    \vspace{-0.4 cm}
\end{figure}

Finally, the receiver operating characteristic (ROC) curves are plotted in Fig.~\ref{fig:fig5} to illustrate the target detection performance.
It can be concluded that the proposed RIS-ISAC system exhibits significantly better performance for fluctuating target detection owing to our proposed joint beamforming design for RIS-empowered multipath exploitation.

\section{Conclusion}
In this letter, we investigated a typical RIS-assisted ISAC scenario where the BS serves several downlink communication users and detects a fluctuating target with the exploitation of a controllable multipath.
We jointly designed transmit beamforming and RIS passive beamforming to maximize the expectation of radar SNR under the constraints of communication SINR, the total transmit power budget, and the RIS reflection coefficients.
Numerical results demonstrated the advantages of exploiting spatial diversity for fluctuating target detection in RIS-aided ISAC systems.
Motivated by this initial work, we will consider utilizing other more advanced RIS architectures for multipath exploitation in ISAC scenarios, such as active RIS \cite{K. Zhi_CL_2022} and stacked intelligent metamaterials (SIM) \cite{J. An_WC}.

\end{document}